# Introducing Artificial Intelligence Agents to the Empirical Measurement of Design Properties for Aspect Oriented Software Development

Senthil Velan S
*Department of Computer Science and Engineering*
*Amity University*
Dubai, UAE
svsugana@gmail.com, svelan@amityuniversity.ae

*Abstract*—The proponents of Aspect Oriented Software Development (*AOSD*) methodology have done a tremendous amount of work to bring out the positive effects of its adoption using quantitative assessment. A structured assessment of the methodology requires a well-defined quality model. In this paper, an AI agent based quality model has been proposed to evaluate the effect of aspectization. The model has been applied on already existing and equivalent versions of object oriented and aspect oriented case study application, university automation software. Specific metrics for the software design properties have been measured using *AI* agents for the different versions and were used to infer upon the effect on quality. Based on the initial measurement, it was found that aspectization has positively improved all the three quality attributes defined in the quality model. The attributes of evolution and reusability showed significant improvement in quality due to the effect of aspectization.

## I. Introduction

Aspect Oriented Software Development (AOSD) [1] is a twenty year old software development methodology focusing on the clear separation of core and cross-cutting functionalities or concerns in a software. By doing so the proponents of this unique methodology profess that the modularity of the software will be positively improved and thereby reducing its maintenance cost. In order to understand the effects of aspectizing an Object Oriented application, sufficient number of empirical studies need to available in the literature. Also, the effect on the overall quality of the software must be analyzed and understood to make a decision on adopting this relatively new method of software development.

In the literature [2–7] a number of authors have done multiple work in measuring the effect of aspectizing a non-*AO* software across its versions. Since the quality of a software cannot be directly measured, the focus of all their work has been on defining new set of metrics for quantifying the software design properties. The design properties of cohesion, coupling, inheritance, ripple effect, etc… have been considered for measurement. Unique set of metrics have been defined and directly applied on *OO* and their equivalent *AO* versions.

Artificial Intelligence (*AI*) is a branch of computer science involving the development of a theoretical framework for intelligence in machines by simulating the thinking framework of human beings. *AI* is one of the fastest growing area since intelligence is acquired by processing the huge amount of data currently available across different repositories. Knowledge can be the result of processing humongous amount of data stored in the repositories. This knowledge can be used to train the machines for providing effective prediction of results.

*AI* can be applied in measurements by training the machines to learn about the data acquired over a period of time. *AI* agents are software entities that can process the measured data and can compute the results based on the trained values. In the empirical study on aspectization, the *AI* agents can be assigned or trained based on the data obtained from the different and equivalent versions of the *OO* and *AO* applications. This can be used for developing an effective prediction model in the study of applying *AOP* across different versions of software applications.

The empirical study on the effect of aspectization using *AI* agents requires a process work flow. A brief work flow of the process elements is shown in Figure 1. The first process is the identification of core design properties which can be directly measured using relevant measurements. The subsequent step is to define a set of metrics that can used for quantifying the design properties identified in the previous step. The third step in the process is to measure the values of the metrics and to learn about the design properties using *AI* agents. The final step in the process is to infer on the effect of aspectization on the higher level quality attributes.

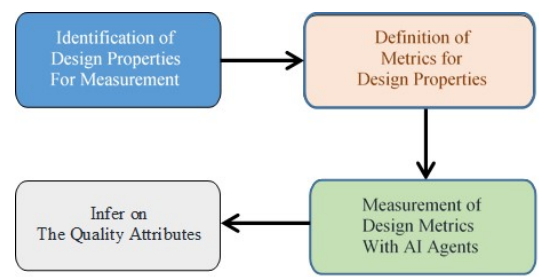

Fig. 1. Workflow of AI Agent based Quality Model

The rest of the paper is organized as follows: Section II elaborates on the related work available in the literature addressing the different measurement available for the measurement of design properties in *AOSD*. The second part of this section explain the need for the introduction of Artificial Intelligence (*AI*) agents in measurement to infer on the quality of the software. Section III explains the proposed quality model with the introduction of *AI* agents during measurement. The different set of metrics used for the measurement is briefly explained in Section IV. The existing case study which we have developed in our previous work [8-10] and on which the model has been applied is explained in Section V. Later, Section VI discusses the results obtained by applying the quality model. Finally, concluding remarks are written in Section VII, with pointers to the possible future work.

## II. EXISTIING WORK

### A. Empirical Evaluation of AOSD

A two decade old methodology like *AOSD* requires a number of empirical studies to understand the effect of aspectizing *OO* applications over its versions. Deepiga et al [8] have done empirical investigation on versions of *SOA* based *Java* and *AspectJ* application over its versions. Application of *AI* based techniques in the measurement can improve the analyzing the effect of aspectizing the case study application.

Measuring the impact of core software design properties have also been attempted by Parthipan et al [9] and Vinobha et al [10]. The design properties considered by the authors were complexity and inheritance respectively. Effects of aspectization have been inferred based only on directly measuring the *OO* and equivalent *AO* versions. An *AI* agent based empirical evaluation will improve the quality of inferences while studying the effect of using *AOSD*.

### B. Application of AI in Empirical Evaluation of Systems

Several authors have studied the application of Artificial Intelligence in empirical evaluation and measurement of software. In one of the first work done by Alippi et al [11], a survey has been done on the effectiveness of applying *AI* in the field of instrumentation and measurements. A similar study in the measurement and inference of applying *AOP* technique to *OO* software has not been attempted and found in the literature.

Amigoni et al [12] have studied the role played by the field of *AI* in the development and use of measurement based systems. They have also explained the motivations behind adopting multi-agent systems as a modern measurement system. A similar agent based measurement system for *AOP* will enhance the awareness of aspectizing an *OO* based application.

In a work done by Neelamdhab et al [13], an attempt has been made to define a model focusing on applying artificial intelligence in evolutionary computing for the estimation of reusability of software. The model has been applied on software developed using the *OO* methodology and a similar model for *AO* will help to generate more knowledge on the impact of using *AOP* for software development.

## III. PROPOSED AI AGENT BASED AO QUALITY MODEL

A unique property of the software is that its quality cannot be directly measured using a quantification tool. Hence, there is a need for a quality model to apply on the software under consideration for measurement. This quality model can be defined using multiple layers with well-defined relationship between the layers. The proposed quality model to infer upon the effect of aspectization is shown in Figure 2.

### A. LAYER 1 – Design Properties

The first layer of the proposed quality model is the list of core design properties of both *AO* and *OO* software considered for the study on quality characteristics. The core design properties considered in the quality model are Inheritance, Cohesion, Coupling, Complexity and Ripple Effect. Inheritance is an important property that could play a strong role in the extension of existing modular elements. Similarly, cohesion and coupling provide the modular strength and inter-dependence among modules of the software. Reducing the complexity of the modular elements can in turn improve the ability to understand the functionalities defined in the software. Ripple effect of changes made during the evolution could directly affect the maintenance of software modules.

### B. LAYER 2 – Proposed Metrics

Measurement is the core part of the empirical studies. Since, the quality of a software could not be directly measured, these metrics will help in measuring the software design properties defined in Layer 1 of the quality model. Specific metrics are defined while accurately capturing the relevant design properties. Some of the metrics are defined in our previous set of research work [4, 9, 10]. The measured values can further be analyzed in order to generate inferences about the effect of aspectization.

### C. LAYER 3 – AI Measurement Agents

Layer 3 of the quality model involves the *AI* agents programmed to both measure and learn about the metrics defined in the previous layer. Learning happens while measuring the different incremental versions of *OO* and their equivalent *AO* versions. Changes in the values of different metrics can be used by the *AI* agents to infer on the effect of aspectizing the *OO* versions of the case study application.

### D. LAYER 4 – Quality Attributes

The fourth layer of the quality model defines three high level quality attributes namely reusability, maintainability and evolution. Reusability is the ability to define and inherit the different modules of a software, while, evolution focuses on the inclusion or removal of functionalities over versions of the software. Similarly, maintainability is another core quality attribute that deals with preventive, corrective, adaptive and perfective capabilities over versions of the software.

## IV. METRICS FOR THE DESIGN PROPERTIES

In the proposed *AI* Agent based Quality Model, there are five important design properties that have been included in Layer 1, namely, Inheritance, Cohesion, Coupling, Complexity and Ripple Effect. The design properties have been quantified using focused set of metrics which has been defined extensively in our previous work [4, 9, 10]. The list of metrics have been defined in the Table I, II, III and IV.

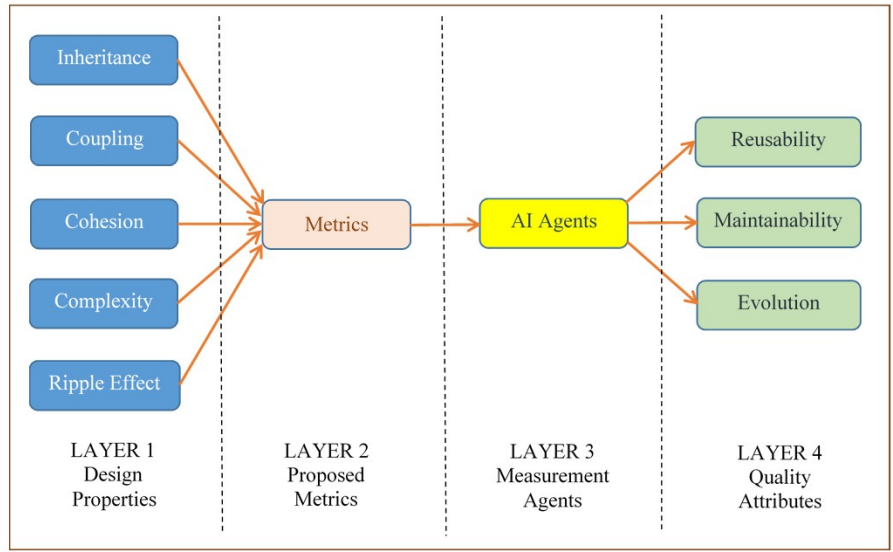

Fig. 2. Components of the proposed AI Agent based AO Quality Model

TABLE I. METRICS FOR INHERITANCE

| S. No. | Name of the Metric | Symbolic Notation |
|---|---|---|
| 1 | Advice Inheritance Factor | AdIF |
| 2 | Pointcut Inheritance Factor | PIF |
| 3 | Attribute Inheritance Factor | AttIF |
| 4 | Aspect Inheritance Factor | AIF |
| 5 | Class Method Inheritance Factor | CMIF |
| 6 | Class Inheritance Factor | CIF |

TABLE II. METRICS FOR COHESION AND COUPLING

| S. No. | Name of the Metric | Symbolic Notation |
|---|---|---|
| 1 | Cohesion Index | CoI |
| 2 | Coupling Index | CuI |

TABLE III. METRICS FOR COMPLEXITY

| S. No. | Name of the Metric | Symbolic Notation |
|---|---|---|
| 1 | Weighted Pointcuts in Aspects | WPA |
| 2 | Weighted Advices in Aspects | WAA |
| 3 | Weighted Join Points in Classes and Aspects | WJPCA |
| 4 | Weighted Methods in Classes and Aspects | WMCA |
| 5 | Weighted Attributes in Classes and Aspects | WACA |

TABLE IV. METRICS FOR RIPPLE EFFECT

| S. No. | Name of the Metric | Symbolic Notation |
|---|---|---|
| 1 | Aspect Change Propagation Reachability | ACPR |
| 2 | Pointcut Change Propagation Reachability | PCPR |
| 3 | Advice Change Propagation Reachability | ADCPR |
| 4 | Class Change Propagation Reachability | CCPR |
| 5 | Method Change Propagation Reachability | MCPR |
| 6 | Complexity of Weaving | CW |
| 7 | Complexity of Control Flow | CCF |

The new set of metrics for measuring the use of inheritance is given in Table I. These metrics capture the occurrence of inheritance found in the different encapsulated units of *OO* and *AO* software. Similarly, a new pair of metrics have been defined for the design properties of cohesion and coupling and are listed in Table II. Complexity is another important attribute of interest and can again be found in the different units of encapsulation. These metrics are listed in Table III. Ideally, the effects of changes made to the units a software need to be constrained within boundaries of the changed modules. If the ripple effect of the changes affects across more number of modules, then maintenance cost will increase over a course of time. The ripple effect of changes is captured with metrics listed in Table IV.

## V. CASE STUDY APPLICATION

The proposed model in this paper requires an application with multiple versions which have been developed using the concepts of *OO* an *AO* methodology. In order to test our work an *SOA* application, consisting of an appreciable number of modelled core and cross-cutting concerns, is indeed needed to understand the impact of the different design attributes in both *OO* and *AO* software. An existing application that we have already developed in our previous work [8-10] namely, the University Automation System (*UAS*) Application which has been developed to automate the typical operations of a university, has been considered as a case study, since it is modelled with a good number of tangled and scattered concerns. Because of these properties, the case study, *UAS* has been identified to understand the effect of aspectization in the available versions of *AO* software. Only core concerns are modelled in the *UAS AJ 1.0* version. The cross-cutting functionality of logging and persistence have been included in version, *UAS AJ 1.1*. The *Login web service* modelled in both the staff and the student have the encapsulated unit of logging aspect. This web service is needed so that it can record both the details of the logged-in user and the timing information. The *Persistence* aspect is encapsulated in to the *Register* web service of both categories of the users. This is needed to store the registration details in the database and persist the obtained registration credentials. The functionality in the record *events()* available in *file log* are inherited by the *audit log* and *event log*, which are defined in the *logging* aspect. The translation of SQL is also inherited from the *Persistence* aspect. Database connections are defined in the Abstract *Persistence* aspect. The operations of insert and update operations for a database are define in the set of concrete aspects. Fig 3 depicts the structural information of the case study application.

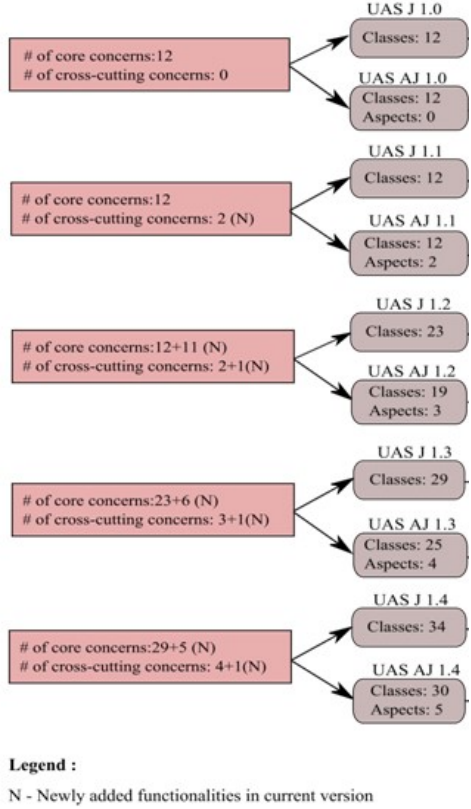

Fig. 3. Structural Information about the versions of *UAS*

Version *UAS AJ 1.2* is modeled with one additional cross-cutting concern namely, security. Because, the login concern is scattered and modeled for all the users, the authentication functionality is separated and modeled as a security aspect in *AOP*. The observer pattern is modeled as a cross-cutting concern in *UAS AJ 1.3* version of the case study. When the result is updated, the updation will be notified to the student by applying the concept of observer pattern. The observer Register observer is modeled as an abstract aspect and the observer Notify observer is modeled as a concrete aspect. The cross-cutting concern of exception handling is modeled in version *UAS AJ 1.4*. The traditional *OOP* implementation the exception handling concern is tangled with the primary concern. Due to the implicit coupling between the tangling of concerns, problems can occur during the maintenance and evolution. In order to avoid such a scenario, exception handle code in *UAS* can be carved out and modeled as an aspect so that the thrown exceptions in *UAS*, can be effectively handled in this *UAS* version. The different types of exceptions, namely, IO exception, class not found exception and runtime exception inherits the functionality of *displayexception()* from abstract exception handling aspect.

## VI. RESULTS AND DISCUSSION

The *AI* based *AO* Quality Model was applied on the different versions of the *UAS* case study application. The initial measurements taken were found to throw several insights in the effect of aspectization. Since more than one design property has been measured and using the well-defined set of metrics, the discussion will focus on the individual effect and further on the overall effect.

The different artifacts that are affected due to the changes in the *Java* and *AspectJ* versions of *UAS* have been measured and the Change Propagation Reachability metric values were calculated using the *AI* agents. The measurements have been analyzed to understand the ripple effect in both *OO* and *AO* versions. The analysis shows that the ripple effects have reduced by 165.6% in the *AspectJ* versions of *UAS* compared to the equivalent *Java* versions. The measurements also indicate that the ripple effects in methods and advices are less in *AspectJ* versions. This reduction of ripple effect in the *AspectJ* versions positively improves the maintainability on an average of 54%, compared to the equivalent *Java* versions.

Based on the measurements for cohesion and coupling, it was found that *AO* versions of the case study application have shown higher values of cohesion index (an increase of 7.6%) than the equivalent *OO* versions. Similarly, the coupling indexes of *AO* versions were lower (less by 9.3%) than their equivalent *OO* versions. Aspectization of the *OO* versions of *UAS* has improved their maintainability, an average increase of 8.4%, due to increase in modularity of artifacts.

The measurements for the complexity performed and it was found that the complexity index of the *AO* versions were comparatively low, an average decrease of 33.6% than their equivalent *OO* versions. Hence, the artifacts in the *AspectJ* implementations of *UAS* were better maintainable (increased by 33.6%) compared to the equivalent *Java* implementations.

The measurements of inheritance of artifacts, both classes and aspects and its internal constructs, showed an improvement in its reusability (an increase of 117.8%). Multi-level inherited functionalities such as *logging*, *persistence* and *authentication* have been scattered in *Java* versions of *UAS* which leads to duplication of reusable elements. Whereas, in aspectized versions, the same functionalities have been neatly modularized thereby achieving a better form of reusability, reflected by the increase in values of the respective metrics over *AspectJ* versions. Based on the measurement, inheritance is increased by 69.9% in *AO* elements which are based on higher modular *AO* components.

Finally based on the proposed quality model, it was found that *AOSD* was able to (i) positively improve by 117.8%, the reusability of modular units either encapsulated as aspects or classes, (ii) reduce the complexity (33.6%) and subsequently improve the maintainability (an overall increase of 54%), (iii) increase the cohesion (7.6%) of the encapsulated units leading to improved reusability (overall increase of 63.1%) of aspects and classes, and (iv) improve the evolution by 165.6%.

VII. CONCLUSION AND FUTURE WORK

Understanding the effect of aspectization will enable the software designers to make effective decisions on adopting *AO* methodology for the development of software. In this work, an *AI* Agent based *AO* Quality model has been proposed which has been composed with four layers. A case study application which was developed by our previous work has been used to apply the proposed Quality Model. Based on the obtained results it was found that aspectization has positively improved the quality of *AO* software over its version. Some of the possible directions for future research are briefly stated in the following paragraph.

Analyzing the trade-offs between various quality attributes, when *AOSD* is practiced in the software industry can also be attempted as an extension. In order to generate more general inferences on the use of *AOSD*, the evaluation models can be applied to more case study applications belonging to different domains. This can also provide an insight on domain-specific, functional and non-functional specific aspectization of *OO* software. Best practices in aspectizing a *OO* software can also be deduced by refactoring different functionalities of *OO* software. This can provide guidelines for learners to choose appropriate refactoring while aspectizing an existing software using *AOSD*. Further, *AI* agents can acquire more knowledge and prediction capabilities which will be the primary focus of the future research.

REFERENCES

[1] Kiczales G., Lamping J., Mendhekar A., Maeda C., Lopes C. V., Loingtier J. M., and Irwin J., "Aspect-Oriented Programming," In *European Conference on Object Oriented Programming*, 1997, pp. 220–242.

[2] Zhao, Jianjun, and Baowen Xu, "Measuring Aspect Cohesion," In *International Conference on Fundamental Approaches to Software Engineering*, pp. 54-68. Springer, Berlin, Heidelberg, 2004

[3] Zhao J., "Measuring Coupling in Aspect-Oriented Systems," in *Information Processing Society of Japan* (*IPSJ*), 2004, pp. 14–16.

[4] Senthil Velan S., Chitra Babu, and Madhumitha R., "A Quantitative Evaluation of Change Impact Reachability and Complexity across Versions of Aspect Oriented Software," In *The International Arab Journal of Information Technology*, Vol. 14, No. 1, January 2017, pp. 41–52.

[5] Pataki N., Sipos A., and Porkolab Z., "Measuring the complexity of aspect-oriented programs with multiparadigm metric," In *Proceedings of the 10th ECOOP Workshop on Quantitative Approaches in ObjectOriented Software Engineering* (*QAOOSE 2006*), 2006, pp. 1-10.

[6] Sheela, G. Arockia Sahaya, and A. Aloysius., "Aspect Oriented Programming-Cognitive Complexity Metric Analysis Tool," *International Journal of Scientific Research in Computer Science, Engineering and Information Technology*, Vol. 3, Iss. 1, 2018, pp. 480-486.

[7] Ceccato M. and Tonella P., "Measuring the Effects of Software Aspectization," *1st Workshop on Aspect Reverse Engineering* (*WARE*), 2004.

[8] Deepiga A S, Senthil Velan S and C. Babu, "Empirical investigation of introducing Aspect Oriented Programming across versions of an SOA application," *2014 IEEE International Conference on Advanced Communications, Control and Computing Technologies*, Ramanathapuram, 2014, pp. 1732-1739.

[9] Parthipan S, Senthil Velan S and C. Babu, "Design level metrics to measure the complexity across versions of AO software," *2014 IEEE International Conference on Advanced Communications, Control and Computing Technologies*, Ramanathapuram, 2014, pp. 1708-1714.

[10] Vinobha A, Senthil Velan S and C. Babu, "Evaluation of reusability in Aspect Oriented Software using inheritance metrics," *2014 IEEE International Conference on Advanced Communications, Control and Computing Technologies*, Ramanathapuram, 2014, pp. 1715-1722.

[11] C. Alippi, A. Ferrero and V. Piuri, "Artificial intelligence for instruments and measurement applications," in *IEEE Instrumentation & Measurement Magazine*, vol. 1, no. 2, June 1998, pp. 9-17.

[12] F. Amigoni, A. Brandolini, G. D'Antona, R. Ottoboni and M. Somalvico, "Artificial intelligence in science of measurements: from measurement instruments to perceptive agencies," in *IEEE Transactions on Instrumentation and Measurement*, vol. 52, no. 3, June 2003, pp. 716-723.

[13] Neelamdhab Padhy, R.P. Singh, Suresh Chandra Satapathy, "Software Reusability Metrics Estimation: Algorithms, Models and Optimization Techniques," *Computers & Electrical Engineering*, Volume 69, 2018, pp. 653-668.